\documentclass[twocolumn,showpacs,preprintnumbers,amsmath,amssymb]{revtex4}
\usepackage{graphicx}% Include figure files
\usepackage{dcolumn}% Align table columns on decimal point
\usepackage{bm}% bold math
\usepackage{mathrsfs}
\begin{document}

% Use /ltx/revtex4/sample/apssamp.tex as a template
%\preprint{APS/123-QED}

\title{Galaxy clusters and the nature of Machian strings}

\author{David W. Essex}
 \email{D.W.Essex@damtp.cam.ac.uk}
 \affiliation{Clare College, Trinity Lane, Cambridge, CB2 1TL, England}
% \affiliation{Centre for Mathematical Sciences, Wilberforce Road, Cambridge, CB3 0WA, England}
% \homepage{http://www.homepage}
\date{\today}

\begin{abstract}
 The model of Machian space quanta is applied to the dark matter problem in galaxy clusters. 
%It is known that Newtonian gravity fails to provide sufficient gravitational acceleration to hold a cluster together. 
%Although very successful for individual galaxies, MOND also fails in a galaxy cluster.
Machian space quanta are able to solve the missing mass problem
%, in both galaxies and galaxy clusters, 
if all the mass of a space quantum is assumed to be in the Machian strings.

%\vspace*{0.2cm}

% Valid PACS numbers may be entered using the \verb+\pacs{#1}+ command.
\end{abstract}
\maketitle

\section{Introduction}\label{intro}
 A galaxy cluster consists of a few thousand galaxies held together by gravity. Early measurements of the velocities of individual galaxies in the Coma cluster~\cite{zwicky} suggested that the amount of matter needed to hold the galaxies together, assuming Newtonian gravity, is hundreds of times larger than the amount of visible matter in the galaxies. X-ray studies later revealed the presence of large amounts of intergalactic gas, with a mass some ten times larger than the mass in the galaxies~\cite{white}, but the total baryonic mass in the cluster is still an order of magnitude smaller than required by Newtonian gravity.
%Either clusters contain a large amount of mysterious dark matter or Newtonian gravity is incomplete.

%The simplest approach to modified gravity is to postulate a modification to
It is well known that the missing mass problem for a single galaxy may be solved by postulating a simple modification to Newton's law of gravity~\cite{milgrom,mcgaugh2}, known as MOND, in which the gravitational acceleration is changed from a $1/r^2$ law to a $1/r$ law when the acceleration is smaller than than a certain critical acceleration $a_0\approx 1.2\times 10^{-10}$m/s$^2$. Unfortunately, MOND does not solve the missing mass problem in galaxy clusters. In the Coma cluster, for example, the gravitational acceleration needed to hold the cluster together is of order $a_0$ but the gravitational acceleration in MOND has a maximum of about $0.3a_0$, which is three times too small.

In systems of interacting galaxy clusters, such as the famous bullet cluster~\cite{clowe}, the centres of the galaxy and gas distributions may become separated and a more fundamental problem is then revealed. In a modified gravity theory such as MOND, which simply gives an enhancement of the existing Newtonian gravitational field, the gravitational mass distribution is necessarily centred on the dominant mass component in the system, namely the gas. However, gravitational lensing studies clearly show that the gravitational mass distribution is actually centred on the galaxies. The bullet cluster system cannot be understood using a simple modification of Newtonian gravity and is often cited as definitive evidence for the existence of dark matter.

%Brownstein and Moffat claim that their theory of modified gravity can explain X-ray cluster masses without dark matter and they also claim that it can account for the observations of the Bullet Cluster.
%  Observations of the Bullet Cluster are the strongest evidence for the existence of dark matter;[8][9][10] however, Brownstein and Moffat[11] have shown that their modified gravity theory can also account for the properties of the cluster.

In a recent paper~\cite{paper1}, a new approach to modified gravity was suggested based on the idea that elementary particles are not point particles but are, in fact, Machian space quanta with the size of the observable universe. The Machian space quantum of a massive particle appears to be point-like, since it has a point-like charged centre,
but the mass of the particle is actually distributed throughout the Machian strings. Newtonian gravity is due to the energy in the direct strings joining the centres of two masses and dark matter effects are attributed to the interactions between the Machian strings joining their centres to the centres of all the other masses in the Universe. The purpose of the present paper is to show that the Machian string model can solve the missing mass problem, without dark matter, in both galaxies and galaxy clusters.

\begin{figure}[h]
\vspace*{0.5cm}
\includegraphics[height=6cm,width=7cm]{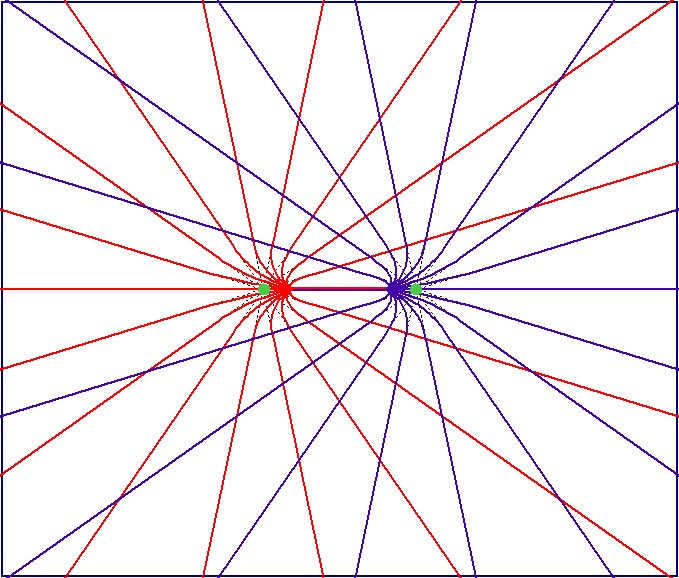}
\caption{\label{sbullet} Schematic diagram of the Machian strings of the gas particles in the bullet cluster. The centres of the galaxy distributions are shown by the green dots. The Machian strings of the gas particles in the clusters on the left and right are shown in red and blue, respectively, and the dashed black lines show their positions before the gas centres were displaced. The Machian strings of the gas particles are still centred on the galaxies, even when the gas particle centres are displaced inwards. The gravitational lensing is therefore centred on the galaxies, as observed, even though there is much more mass in the gas.}
\end{figure}

To explain the experimental data for galaxy clusters using the Machian string model, two new concepts are required. Firstly, to obtain a sufficiently large additional acceleration, it is necessary to assume that {\it all} the mass of a Machian space quantum is in the Machian strings. Secondly, it is necessary to consider the effect of alignment of the direct strings. The new model is discussed in Section~\ref{msm} and applied to the Coma cluster in Section~\ref{coma}.

The model of Machian space quanta can easily account for the bullet cluster observations, at least qualitatively, by considering the Machian strings of the gas particles. During the collision between the two galaxy clusters in the system, the gas particle centres were displaced relative to the galaxy centres by the electromagnetic interaction, as illustrated schematically in Figure~\ref{sbullet}. Since Machian strings are not charged, the entire length of the Machian strings was unaffected apart from small sections at the ends where the strings join on to the centres of the gas particles. The observational result that the gravitational lensing is centred on the galaxies is then easy to understand because most of the lensing effect comes from the interaction with the Machian strings of the gas. Detailed calculations for the bullet cluster are given in Section~\ref{bullet}.

%The precise nature of the Machian strings is not known so it is not possible at present to derive the interaction between Machian strings from first principles. The best we can do is therefore to try and constrain the interaction using experimental data. In particular, the model must reproduce the MOND phenomenology for individual galaxies and give a larger effect than MOND in the inner region of a galaxy cluster, as well as explaining how the gravitational mass can be offset from the baryonic mass in the bullet cluster. As pointed out in~\cite{paper1}, it is also important to check that the model does not give an unacceptably large additional acceleration in the Solar System. {\it Can we give here an order of magnitude argument to show that the additional acceleration in a cluster has to be at least $a_0$\,? Then we can discuss the new assumption that all the mass is in the strings.}

\section{The Machian string model}\label{msm}

\subsection{The nature of Machian strings}\label{nature}
The analysis of the Coma cluster in Section~\ref{coma} below shows that the Newtonian gravitational acceleration in the cluster is about $0.1a_0$ whereas the acceleration needed to hold the cluster together is about $a_0$. The maximum additional acceleration in the model considered in~\cite{paper1} is only about $0.2a_0$, which is clearly too small. However, the calculations were based on the assumption that only about $8\%$ of the total mass of a particle is in the strings. If the fraction of mass in the strings is actually higher then the strings would have higher energies and would exert greater forces on the centres of a test particle, leading to a larger additional acceleration. The simplest possibility from a conceptual point of view is that {\it all} the mass of a space quantum is in the strings, with none at the centres. The calculations in the present paper are based on the assumption that all the mass of a space quantum is indeed in the strings.

\begin{figure}[h]
\vspace*{0.5cm}
\includegraphics[height=4.8cm,width=7.5cm]{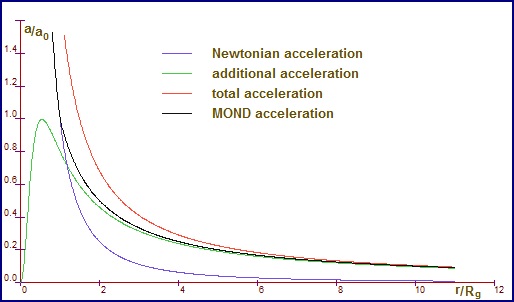}
\caption{\label{newfig3} The additional acceleration on a test particle, due to the interaction with the Machian strings of a point mass $M$, for the interaction function considered in Appendix~\ref{appfu}. The acceleration is plotted in units of $a_0$ as a function of $r/R_g$, where $R_g$ is the distance from the mass $M$ at which the Newtonian gravitational acceleration is equal to $a_0$. All the mass of a space quantum is now assumed to be in the strings, in contrast to the model discussed in~\cite{paper1}, so the maximum additional acceleration is much larger.}
\end{figure}

With much larger additional accelerations from the interactions between Machian strings, it is important to reconsider the constraint imposed by the absence of dark matter effects in the Solar System. As noted in~\cite{paper1}, the experimental data puts a limit on any additional gravitational acceleration in the Solar System at the orbit of Saturn of about $7.5\times 10^{-5}\,a_0$. A specific model, defined by specifying a particular form for the interaction between Machian strings, that agrees with MOND in a galaxy and gives an additional acceleration of order $a_0$ in a galaxy cluster while still respecting the Solar System constraint is described in Appendix~\ref{appfu}. The resulting additional acceleration due to a point mass $M$ is shown in Figure~\ref{newfig3}, which is an updated version of Figure~3 in~\cite{paper1}.

It is important to note that Figure~\ref{newfig3} gives the additional acceleration in a system of any total mass whatsoever, whether the system is a planet, the Sun or a galaxy, provided the system may be considered to be a point mass. In fact, Figure~\ref{newfig3} applies even more generally because, even for an extended system such as a galaxy cluster, the Machian strings are all pulled inwards to the centre of the system by their mutual interaction. Thus, even though the particle centres may be highly delocalised, the distribution of Machian strings around them is almost the same as if the particle centres were all at a point. Since the additional acceleration on a test mass is determined by its interaction with the Machian strings, the additional acceleration is still given by Figure~\ref{newfig3}. The only variable is the length scale $R_g$ which, for a system of total mass $M$, is given by~\cite{paper1} $R_g= 0.32\, R_U\sqrt{M/M_U}$.

\subsection{The alignment of direct strings}\label{align}
A new effect that appears when the mass distribution is both delocalised and very massive is the alignment of direct strings. The energy in the direct strings joining the centres of a mass $M$ and a test mass $m$ is responsible for the usual Newtonian gravitational interaction, as described in~\cite{paper1}. When all the centres of $M$ are at a point, or when the test mass is sufficiently distant, all the direct strings lie along the line of centres and the total force exerted by the strings is the Newtonian gravitational force, $GMm/r^2$. Now consider the direct strings joining a test mass to the centres of a galaxy cluster. Just as the interaction of the Machian strings of the cluster with each other causes the Machian strings to be pulled inwards, the interaction between the Machian strings of the cluster and the direct strings joining the cluster and the test mass causes the direct strings to be pulled towards the line of centres. The total force exerted by the direct strings is then {\it larger} than the Newtonian gravitational force, because the forces in the strings all act in the same direction. The effect of direct string alignment on galaxy rotation curves is discussed in Appendix~\ref{appalign}.

\section{The Coma cluster}\label{coma}

\subsection{The missing mass problem}\label{exp}
Measurements of the X-ray emission from the Coma cluster~\cite{briel} show that the gas density has the form
\begin{eqnarray}\label{cdens}
 \rho(r)\,=\,\rho_0\Big(1 + \frac{r^2}{r_c^2}\Big)^{\!-3\beta/2}\,,
\end{eqnarray}
where $\rho_0$ is the central mass density, $r_c= 0.42/h_{50}$\,Mpc is the core radius and $\beta=0.75$. Taking $h_{50}= 1.4$, corresponding to $H= 70$\,km/s/Mpc, gives $r_c= 300\,$kpc. The central electron density is $n_e=2.9 \times 10^{-3}\sqrt{h_{50}}$\,cm$^{-3}$. For an ionised gas containing hydrogen and $10\%$ helium, the corresponding central mass density is $\rho_0= 6.7\times 10^{-24}$\,kg/m$^3 =1.0\times 10^5\,M_{\odot}$/kpc$^3$.

\begin{figure}[h]
\vspace*{0.5cm}
\includegraphics[height=5.5cm,width=7.5cm]{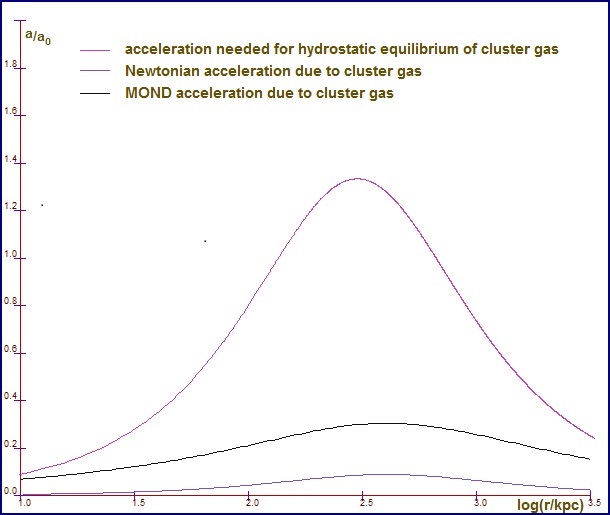}
\caption{\label{cluster4} The acceleration needed to keep the gas in hydrostatic equilibrium in the Coma cluster compared to the Newtonian and MOND gravitational accelerations. The acceleration is plotted in units of $a_0$ as a function of $\log r$, where $r$ is the distance from the centre of the cluster in kpc.}
\end{figure}

The inward acceleration required to keep the gas in hydrostatic equilibrium is given by
\begin{eqnarray}\label{a}
 a(r)\,=\,-\frac{1}{\rho(r)}\frac{dp(r)}{dr}\,,
\end{eqnarray}
where $p(r)$ is the gas pressure. The ideal gas law gives $p= \rho kT/m$ where the average particle mass, $m$, is $0.61$ times the proton mass. The gas is approximately isothermal, with temperature $kT= 8.6$\,keV~\cite{arnaud}, so~(\ref{a}) becomes
\begin{eqnarray}\label{a2}
 a(r)\,=\,-\frac{kT}{m\rho(r)}\frac{d\rho(r)}{dr}
\end{eqnarray}
and substituting for $\rho(r)$ from~(\ref{cdens}) then gives
\begin{eqnarray}\label{a3}
 a(r)\,=\,\frac{3\beta kT}{m}\frac{r}{r^2+r_c^2}\,.
\end{eqnarray}
The constant $3\beta kT/m$ is equal to $3.0\times 10^{12}$\,m$^2$/s$^2$, which may be written as $800a_0$\,kpc, so if $r$ and $r_c$ are in units of kpc then
\begin{eqnarray}\label{a4}
 \frac{a(r)}{a_0}\,=\,\frac{800\,r}{r^2+r_c^2}\,.
\end{eqnarray}
The function~(\ref{a4}), which has a maximum of $1.3$ at $r= r_c$, is plotted in Figure~\ref{cluster4}.

The Newtonian gravitational acceleration corresponding to the gas density $\rho(r)$ is given by $g_N(r)= GM(r)/r^2$, where $M(r)$ is the mass enclosed within a radius $r$, so
\begin{eqnarray}\label{newton}
 g_N(r)\,=\,\frac{4\pi G\rho_0}{r^2}\int_0^r s^2\Big(1 + \frac{s^2}{r_c^2}\Big)^{-3\beta/2}ds\,.
\end{eqnarray}
The constant $4\pi G\rho_0$ is equal to $5.6\times 10^{-33}$\,s$^{-2}$, which may be written as $1.4\times 10^{-3}$\,kpc$^{-1}a_0$. Thus, with distances measured in kpc,
\begin{eqnarray}\label{newton2}
 \frac{g_N(r)}{a_0}\,=\,\frac{1.4\times 10^{-3}}{r^2}\int_0^r s^2\Big(1 + \frac{s^2}{r_c^2}\Big)^{-3\beta/2}ds\,.
\end{eqnarray}
The acceleration~(\ref{newton2}) is also plotted in Figure~\ref{cluster4}. The huge difference between the Newtonian gravitational acceleration and the acceleration needed to keep the cluster gas in hydrostatic equilibrium illustrates the magnitude of the missing mass problem in galaxy clusters.

\subsection{The failure of MOND}\label{mond}
Figure~\ref{cluster4} shows that the required gravitational acceleration in a galaxy cluster is of order $a_0$, whereas the Newtonian gravitational acceleration, $g_N$, is at most $0.1a_0$. The regime $g_N\ll a_0$ is known as the deep MOND regime and the MOND acceleration then has to take the form $g_M=\sqrt{a_0g_N}$ to give a good fit to galaxy rotation curves. Since the largest value of $g_N$ is about $0.1a_0$, the largest value of
$g_M$ is about $0.3a_0$, which is still three times too small. More precisely, the MOND gravitational acceleration, $g_M$, is defined by the equation $g_N= g_M\mu(g_M/a_0)$, where the interpolating function $\mu(x)$ is equal to $1$ when $x\gg 1$ and $x$ when $x\ll 1$. Solving for $g_M$ for the particular choice $\mu(x)= x/\sqrt{1+x^2}$~\cite{milgrom2} gives
\begin{eqnarray}\label{mond2}
 g_M(r)\,=\,g_N(r)\sqrt{\frac{1}{2} + \sqrt{\frac{1}{4} + \Big(\frac{a_0}{g_N(r)}\Big)^2}}\,.
\end{eqnarray}
The acceleration~(\ref{mond2}) corresponding to the Newtonian acceleration~(\ref{newton2}) is shown in Figure~\ref{cluster4}.

%{\it The discussion of the limits of $a_S$ and the form of $f(u)$ uses constraints from galaxies and the Solar system and doesn't involve clusters at all, so it should be discussed in a Section before the discussion of cluster data with details in an Appendix. In the main section we should give an updated version on Fig 3 of the first paper. On the other hand, the form of $f(u)$ was chosen to ensure a maximum acceleration of at least $a_0$, and indeed this was the motivation for making $\kappa= 1$, but maybe it's sufficient in the Introduction to just say that a maximum acceleration of at least $a_0$ will be needed.}

%If $R_g$ denotes the radius of the visible edge of the galaxy, the gravitational acceleration outside the galaxy is therefore $a_0R_g/r$. The ratio $u$ may be written in the form $u= (GM/r^2)/(0.1a_0)$. The Newtonian acceleration is also $a_0$ at the edge of the galaxy, from which it follows that $R_g= 0.32 R_U\sqrt{M/M_U}$ and hence that $\rho= 0.32\, r/R_g$. {\it Need to say this in a simpler way\,!}

\subsection{Machian strings in the Coma cluster}\label{mc}
\begin{figure}[h]
\vspace*{0.5cm}
\includegraphics[height=5.5cm,width=7.5cm]{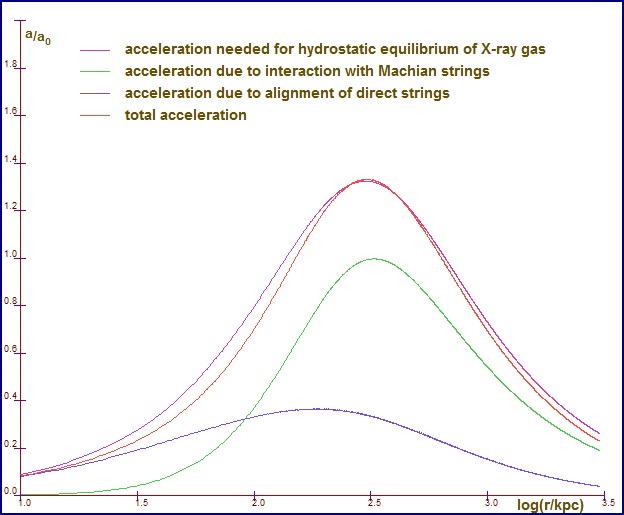}
\caption{\label{acoma} The contributions to the total acceleration on a test particle in the Coma cluster in the Machian string model. The contribution from the Machian strings, shown in green, is obtained by replotting the green curve in Figure~\ref{newfig3} as a function of $\log r$.}
\end{figure}

As discussed in Section~\ref{msm}, all the Machian strings are pulled in to the centre of the cluster and the cluster may be treated as a point mass for the purpose of calculating the additional acceleration due to the Machian strings. The additional acceleration is therefore given by the green curve in Figure~\ref{newfig3}, where the length scale $R_g= 0.32\, R_U\sqrt{M/M_U}$ is determined by the total mass, $M$, of the cluster.
%where the radial coordinate $\rho$ is defined by $r=\rho\sqrt{M/M_U}$ and $M$ is the total mass of the cluster.
 The density~(\ref{cdens}) has to be truncated at some radius $r_t$ to obtain a finite total mass, so the total mass is sensitive to the choice of $r_t$. Observations~\cite{briel} show that $r_t$ is at least $3$\,Mpc and an upper limit can be obtained~\cite{the} by taking the period of circular orbits at the radius $r_t$ as an estimate of the time needed to establish hydrostatic equilibrium. The acceleration~(\ref{a4}) at the radius $r_t\gg r_c$ is approximately $800a_0/r_t$ and the corresponding orbital period is $3.6(r_t$/kpc$)$Myr. The age of the Universe is about $1.4\times 10^4$\,Myr, which suggests that $r_t$ cannot be much larger than about $4$\,Mpc. Taking $r_t= 4\,$Mpc gives a total mass of gas in the Coma cluster of about $2.6\times 10^{14}\,M_{\odot}$. The length scale 
%$R_U\sqrt{M/M_U}$ is then 1920\,kpc, 
$R_g$ is then 614\,kpc, taking $R_U= 3.6\times 10^{26}$\,m and $M_U= 10^{22}\,M_{\odot}$.
%so if $r$ is measured in kpc then $\rho= r/1920$. The corresponding plot of $a_S$ as a function of $r$ is shown in the Figure~\ref{acoma}.
The green curve in Figure~\ref{newfig3} is replotted as a function of $\log r$ in Figure~\ref{acoma}.

The contribution to the additional acceleration from the alignment of direct strings was calculated using the method described in Appendix~\ref{appalign}, by first assuming complete direct string alignment and then including the factor~(\ref{falign}) to ensure that the additional acceleration tends to zero at the centre of the cluster. The choice $r_d= r_c/5$ gives the curve shown in blue in Figure~\ref{acoma} and the total acceleration is seen to give a very good fit to the acceleration required to hold the gas in hydrostatic equilibrium.
% as shown in Figure~\ref{acoma}.

%\begin{figure}[h]
%\vspace*{0.5cm}
%\includegraphics[height=5.5cm,width=7.5cm]{profile.jpg}
%\caption{\label{mass} The mass profiles for the Coma cluster.}
%\end{figure}

\section{Machian strings in the bullet cluster}\label{bullet}
\begin{figure}[h]
\vspace*{0.5cm}
\includegraphics[height=5cm,width=8cm]{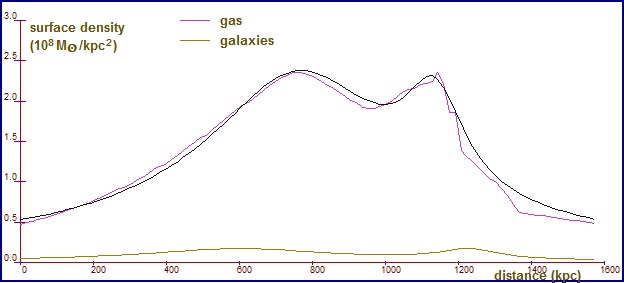}
\caption{\label{surface} The surface density profiles of the gas and galaxy distributions in the bullet cluster from the data given in~\cite{clowe}. The main cluster is on the left and the subcluster is on the right. The $\beta$ model fit to the gas density, shown in black, is discussed in the main text.}
\vspace*{0.5cm}
\includegraphics[height=5cm,width=8cm]{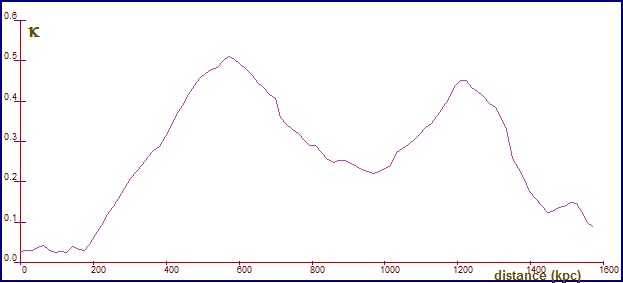}
\caption{\label{kmap} The profile of the $\kappa$ map for the bullet cluster from the data given in~\cite{clowe}.}
\end{figure}

The bullet cluster system consists of two galaxy clusters that collided and passed through each other millions of year ago. During the collision, the electrically charged gas particles were slowed down more than the neutral atoms in the galaxies and the centres of mass of the gas and galaxy distributions became separated. Measuring the intensity of the X-ray emission from different parts of the cluster allows the surface density of gas in the cluster, i.e. the projection of the gas density onto the plane perpendicular to the line of sight, to be determined.
The surface mass densities due to the gas and the galaxies, from the data in~\cite{clowe}, are shown in profile in Figure~\ref{surface}. The system is completely dominated by the gas in the two clusters and it may be seen that the peaks of the gas densities are indeed closer to the centre of the system than the peaks of the galaxy densities.
%The central surface density of gas in the main cluster is $2.1\times 10^8\,M_{\odot}$/kpc$^2$ and the surface density of gas at the centre of the subcluster is $1.7\times 10^8\,M_{\odot}$/kpc$^2$~\cite{clowe2}. The surface density of the gas distribution is shown in profile in Figure~\ref{surface}.
%Figure~\ref{surface} also shows the surface mass density profile of the stars in the galaxy clusters. The maximum surface density of stars is about $1.8\times 10^7\,M_{\odot}$/kpc$^2$, which is over ten times smaller than the density of the gas, and the peaks of the gas density are clearly closer to the centre of the system than the peaks of the galaxy density.

%The total mass of gas in the main cluster is $\sim 3.9\times 10^{14}\,M_{\odot}$ and the total mass of gas in the subcluster is $\sim 2.6\times 10^{13}\,M_{\odot}$. {\it Do we have to cite Brownstein \& Moffat for this analysis\,?
%It may be that we only need the total masses of gas and the positions of the galaxy centres to make our prediction of the kappa map, in which case we don't need to bother describing the beta model fits to the gas densities.}

The gravitational mass density in the bullet cluster can be determined by studying the gravitational lensing effect of the cluster on the images of background galaxies~\cite{clowe}. The lensing effect is described by a quantity known as the lens convergence, $\kappa$, and the $\kappa$ map is proportional to the gravitational surface mass density, as explained in Appendix~\ref{applensing}. 
%For the bullet cluster, $\kappa(x,y)\approx \Sigma_g(x,y)/(3\times 10^9\,M_{\odot}$/kpc$^2)$, where $x$ and $y$ are the position coordinates in the plane of the cluster. 
The $\kappa$ map profile is shown in Figure~\ref{kmap}. Comparison with Figure~\ref{surface} shows that the peaks of the $\kappa$ map, and hence the gravitational mass density, coincide with the peaks of the galaxy density and not with the peaks of the much larger gas density.

To calculate the $\kappa$ map in the Machian string model it is necessary to calculate the effective gravitational surface mass density corresponding to the gas density profile shown in Figure~\ref{surface}. The total gas density may be considered as the sum of two separate gas densities, namely a main cluster gas density associated with the peak on the left in Figure~\ref{surface} and a subcluster gas density associated with the peak on the right. For simplicity, the interaction between the Machian strings of the main cluster and the Machian strings of the subcluster will be ignored, so that the contributions to the $\kappa$ map due to each cluster may be calculated separately and added together.

The gas density shown in Figure~\ref{surface} may be modelled as the sum of two $\beta$ model gas densities, one centred on each peak, of the form~(\ref{cdens}). It is known that a three dimensional gas density of the form~(\ref{cdens}) gives a projected surface mass density of the form
\begin{eqnarray}\label{cdens2}
 \Sigma(r)\,=\,\Sigma_0\Big(1 + \frac{r^2}{r_c^2}\Big)^{\!-(3\beta-1)/2}\,,
\end{eqnarray}
 where $\Sigma_0$ is related to $\rho_0$ by the equation~\cite{brownstein} $\Sigma_0=\sqrt{\pi}\rho_0r_c
 \Gamma(3\beta/2\!-\!1/2)/\Gamma(3\beta/2)$. The fit shown to the profile in Figure~\ref{surface} has $\Sigma_0= 2.3\times 10^8M_{\odot}/$kpc$^2$, $r_c= 270\,$kpc and $\beta= 0.78$ for the main cluster and $\Sigma_0= 1.2\times 10^8M_{\odot}/$kpc$^2$, $r_c= 100\,$kpc and $\beta= 1$ for the subcluster. The corresponding values of $\rho_0$ are $3.3\times 10^5M_{\odot}/$kpc$^3$ and $6.1\times 10^5M_{\odot}/$kpc$^3$, respectively. 

As for the Coma cluster discussed in Section~\ref{mc}, the acceleration due to the Machian strings for a given cluster is given by the green curve in Figure~\ref{newfig3}. The acceleration due to aligned direct strings for each cluster is given by equation~(\ref{adirect2}) in Appendix~\ref{appalign}, and the factor~(\ref{falign}) is again included to ensure that the additional acceleration tends to zero at the centre of the cluster.

Having calculated the additional accelerations towards the two clusters, it remains to calculate the corresponding effective gravitational mass densities. The gravitational mass density, $\rho_g(r)$, corresponding to an acceleration $a(r)$ is defined as the mass density that would give the acceleration $a(r)$ in Newtonian gravity. If $M_g(r)$ is the effective gravitational mass enclosed within a radius $r$ then $a(r)= GM_g(r)/r^2$ and $dM_g(r)/dr= 4\pi r^2\rho_g(r)$, so
\begin{eqnarray}\label{eff}
 \rho_g(r)\,=\,\frac{1}{4\pi Gr^2}\frac{d}{dr}[r^2a(r)]\,.
\end{eqnarray}
If $a(r)$ is in units of $a_0$ and distances are in units of kpc then, in units of $M_{\odot}$/kpc$^3$,
\begin{eqnarray}\label{eff2}
 \rho_g(r)\,=\,\frac{7.1\times 10^7}{r^2}\frac{d}{dr}[r^2a(r)]\,.
\end{eqnarray}
The projected effective gravitational mass density, $\Sigma_g$, is given by
\begin{eqnarray}\label{sig}
 \Sigma_g(r)\,=\,\int \rho_g(\sqrt{r^2 + z^2})~dz
\end{eqnarray}
and dividing by $3\times 10^9\,M_{\odot}$/kpc$^2$, according to equation~(\ref{kappa}), then gives the contribution to the $\kappa$ map.

\begin{figure}[h]
\vspace*{0.5cm}
\includegraphics[height=6cm,width=8cm]{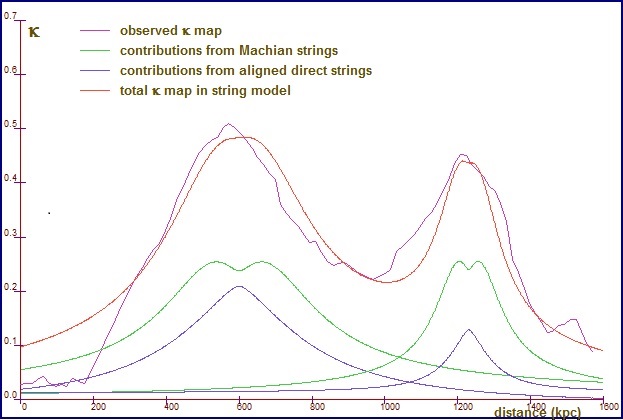}
\caption{\label{kmap2} The observed $\kappa$ map for the bullet cluster from Figure~\ref{kmap} and the contributions to the $\kappa$ map in the Machian string model. The contributions from the Machian strings are shown in green and the contributions from the aligned direct strings are shown in blue.}
\end{figure}

The fit to the data shown in Figure~\ref{kmap2} was obtained for total masses of the main cluster and the subcluster of $1.5\times 10^{14}M_{\odot}$ and $2.7\times 10^{13}M_{\odot}$, respectively, corresponding to $r_t= 1000\,$kpc for the main cluster and $r_t= 5000\,$kpc for the subcluster. The contributions from the aligned direct strings were calculated using $r_d= r_c/2$ for both clusters and the total $\kappa$ map is seen to be in good agreement with the experimental data.

%A smaller total mass gives a smaller value of $R_g$ and a correspondingly smaller width of the effective gravitational mass density. 
% The choice $r_d= r_c/2$, for both clusters, was used to obtain the fit to the data shown in Figure~\ref{kmap2}.   
%The total masses in the main cluster and subcluster gas are estimated to be $3.9\times 10^{14}\,M_{\odot}$ and $2.6\times 10^{13}\,M_{\odot}$, respectively ({\it reference needed}).
%The contributions to the $\kappa$ map are shown in Figure~\ref{kmap2} and the total $\kappa$ map is seen to be in qualitative agreement with the experimental data.

\section{Conclusion}
The model of Machian space quanta is able to solve the missing mass problem in galaxies and galaxy clusters, without dark matter, if all the mass of a space quantum is assumed to be in the Machian strings.

\begin{appendix}
 \section{The interaction between Machian strings}\label{appfu}
The purpose of the calculations below is to show explicitly how to obtain the MOND acceleration in a galaxy and a much larger acceleration in a galaxy cluster without the acceleration in the Solar System becoming larger than the observational limit.

It is convenient to introduce the radial coordinate $\rho$, defined by $r= \rho R_U\sqrt{M/M_U}$, so that the ratio $u= (GM/r^2)/(GM_U/R_U^2)$ of the density of Machian strings around a mass $M$ to the density of background strings is given by $u= 1/\rho^2$. The MOND acceleration required to account for galaxy rotation curves is then $\sqrt{(GM/r^2)a_0}= \sqrt{0.1u}\,a_0= 0.32\,a_0/\rho$. The first requirement of the additional acceleration in the string model is therefore
\begin{eqnarray}\label{as1}
\frac{a(\rho)}{a_0} \sim \frac{0.32}{\rho}\hspace*{0.5cm}{\mbox as}\hspace*{0.5cm}\rho\rightarrow \infty\,.
\end{eqnarray}
As in~\cite{paper1}, the interaction between Machian strings is specified by an interaction function $f(u)$ that gives the fractional increase in mass per unit length in the Machian strings of a test mass due to the presence of a mass $M$. Consider a generalisation of the function $f(u)$ of the form
\begin{eqnarray}\label{fu}
f(u)\,=\,\frac{A(a\sqrt{u} + bu^n)}{1 + a\sqrt{u} + bu^n}\,,
\end{eqnarray}
where $A$, $a$, $b$ and $n$ are parameters to be determined. The function~(\ref{fu}) is approximately $Aa\sqrt{u}$ in the limit $u\ll 1$ and gives an additional acceleration proportional to $\sqrt{u}$ for $u\ll 1$, as required for galaxy rotation curves. In the limit $u\gg 1$, $f(u)$ is approximately $1-1/(bu^n)$. Since a constant increase in mass per unit length of the strings gives no additional forces, the interaction is equivalent to $-1/(bu^n)$ and an additional acceleration proportional to $1/u^n$ is expected. 

In~\cite{paper1}, two ways were given to calculate the additional acceleration. The first was based on a calculation of the total interaction energy and the second was based on numerical calculations of the string paths and the asymmetry of the strings at the centres. 

Consider, first, the total interaction energy. In the limit $\rho >> 1$, the deflection of the Machian strings is very small and may be ignored as far as the calculation of the total energy is concerned. Equations~(I5) and~(I39) of~\cite{paper1} then give, for the total interaction energy in the Machian strings of a test mass $m$,
\begin{eqnarray}\label{int}
 \Delta E(\rho)\,\approx\,\frac{\kappa mc^2}{R_U}\!\int_0^{R_U}\!\!\!\!ds\!\int_0^\pi \!\!\!d\theta\,\sin\theta \,
f[u(|{\bf x}-{\bf x}_M|)]\,,~
\end{eqnarray}
where ${\bf x}_M$ is the position vector of the centre of $M$ relative to the centre of $m$, 
${\bf x}$ and ${\bf x}-{\bf x}_M$ are the positions vector of a point on one of the strings of $m$ relative to the centres of $m$ and $M$, respectively, and $s= |{\bf x}|$. All the mass is now assumed to be in the strings, so $\kappa = 1$. The integral~(\ref{int}) can be evaluated more easily by writing it as an integral over all space. The volume element is $d^3{\bf x}=s^2\sin\theta\, dsd\theta d\phi$, so
\begin{eqnarray}\label{int2}
 \Delta E(\rho)\,\approx\,\frac{mc^2}{2\pi R_U}\int\frac{d^3{\bf x}}{|{\bf x}|^2}
 \,f[u(|{\bf x}-{\bf x}_M|)]
\end{eqnarray}
 and changing the origin to the centre of $M$ then gives
\begin{eqnarray}\label{int3}
 \Delta E(\rho)\,\approx\,\frac{mc^2}{2\pi R_U}\int\frac{d^3{\bf x}}{|{\bf x}+{\bf x}_M|^2}
 \,f[u(|{\bf x}|)]\,.
\end{eqnarray}
 After writing the integration variable in units of $R_U\sqrt{M/M_U}$, equation~(\ref{int3})
becomes 
\begin{eqnarray}\label{int4}
 \Delta E(\rho)\!&\approx&\! mc^2\sqrt{\frac{M}{M_U}}\!\int_0^{\sqrt{\frac{M_U}{M}}}\hspace*{-0.3cm}\int_0^\pi
 \!\!\!\frac{x^2dx\sin\theta d\theta}{x^2 + 2\rho x\cos\theta + \rho^2}\,f[u(x)]\,\nonumber\\
   &=&  \frac{mc^2}{\rho}\sqrt{\frac{M}{M_U}}	\int_0^{\sqrt{\frac{M_U}{M}}}
		\!\!x\,\ln\Big(\frac{x+\rho}{|x-\rho|}\Big)f[u(x)]\,dx\,.\nonumber\\
\end{eqnarray}
For a point mass $M$, $u(x)= 1/x^2$. Now consider the limit $\rho \gg 1$, corresponding to 
$u << 1$. Then $f(u)\approx a\sqrt{u}\approx Aa/x$, so 
\begin{eqnarray}\label{int5}
 && \hspace*{-1.1cm}\Delta E(\rho)\,\approx\,\frac{Aamc^2}{\rho}\sqrt{\frac{M}{M_U}}\int_0^{\sqrt{\frac{M_U}{M}}}\!\!
 \ln\Big(\frac{x+\rho}{|x-\rho|}\Big)\,dx\nonumber\\ 
&& \,= \frac{Aamc^2}{\rho}\sqrt{\frac{M}{M_U}}\Big[(x+\rho)\ln(x+\rho) \vspace*{-0.5cm}\nonumber\\ \noalign{\vskip-3mm}
&& \hspace*{2.5cm}- (x-\rho)\ln(|x-\rho|)\Big]_0^{\sqrt{\frac{M_U}{M}}}\!\!.
\end{eqnarray}
 Expanding in powers of $\rho$ then gives, for $\rho \gg 1$,
\begin{eqnarray}\label{int6}
 \hspace*{-0.4cm}\Delta E(\rho)\,\approx\, 2Aamc^2 \sqrt{\frac{M}{M_U}}\Big[\ln\frac{\sqrt{\frac{M}{M_U}}}{\rho}+1+
 {\mathcal O}(\rho^2)\Big]
\end{eqnarray}
 and the corresponding acceleration, given by~(I42) in~\cite{paper1}, is
\begin{eqnarray}\label{accel}
 a\,=\,\frac{1}{m}\frac{d\Delta E}{dr}&\approx&-\frac{2Aa}{\rho}\frac{c^2}{R_U}\,=\, \frac{5Aa}{\rho}a_0\,.  
\end{eqnarray}
 Comparison with~(\ref{as1}) shows that the parameters $A$ and $a$ are required to satisfy the equation
\begin{eqnarray}\label{aval}
 Aa\,=\,0.064\,.
\end{eqnarray}

The additional acceleration can also be found by calculating the asymmetry in the Machian strings at the centre of $m$ numerically and multiplying by the string tension, as described in Appendices I1 and I2 of~\cite{paper1}. A program was written to calculate the asymmetry in the Machian strings for an interaction function of the form~(\ref{fu}), for different values of the parameters $A$, $a$, $b$ and $n$~\cite{programs}. The numerical calculation confirms the result~(\ref{accel}) in the limit $\rho \gg 1$. For the limit $\rho \ll 1$, the exponent $n$ must be large enough so that the additional acceleration decreases to zero sufficiently rapidly as $\rho$ tends to zero, in order to satisfy the Solar System constraint. The value $n= 3/2$ was found to be sufficient and the corresponding acceleration is given by
\begin{eqnarray}\label{accel2}
 a\,\approx\, \frac{4\rho^3}{b}a_0\,.  
\end{eqnarray}
 According to~\cite{pitjev}, the absence of dark matter effects in the Solar System implies that any additional acceleration is smaller than $7.5\times 10^{-5}\,a_0$ at the radius of Saturn. Taking $u\approx 4.4\times 10^5$ at the radius of Saturn, the corresponding limit on $b$ is $b > 1.8 \times 10^{-4}$. Figure~\ref{newfig3} shows the additional acceleration as a function of $\rho$ when  $A= 0.6$, $a= 0.1$, $b= 0.005$ and $n= 1.5$. With $GM_U/R_Uc^2= 0.1\,a_0$~\cite{paper1}, $\rho$ is related to the ratio $r/R_g$ defined in Figure~\ref{newfig3} by $\rho= 0.32\,r/R_g$.
 \section{Direct string alignment}\label{appalign}
\begin{figure}[h]
\vspace*{0.5cm}
\includegraphics[height=3cm,width=6cm]{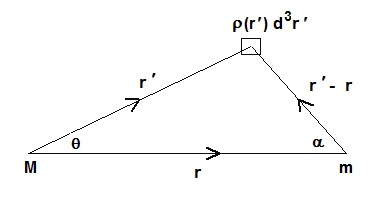}
\caption{\label{spiral} The geometry for calculation of the gravitational acceleration exerted by the direct strings connecting a test mass $m$ to a mass distribution $M$. The direct string joining the mass $m$ to the mass element $\rho({\bf r}^\prime)\,d^3{\bf r}^\prime$ is along the vector ${\bf r}^\prime-{\bf r}$ when the string is straight.}
\end{figure}

Consider the gravitational acceleration exerted on a test mass $m$ at ${\bf r}$ by a mass distribution $\rho({\bf r}^\prime)$, of total mass $M$, due to the direct strings joining them. If the direct strings are all straight then the force between the mass $m$ and a mass element $\rho({\bf r}^\prime)\,d^3{\bf r}^\prime$ of $M$ is in the direction ${\bf r}^\prime-{\bf r}$, as in Newtonian gravity, and the resulting gravitational acceleration is the Newtonian gravitational acceleration given by
\begin{eqnarray}\label{adirect}
a_N(r)\,=\,\int \frac{G\rho({\bf r}^\prime)}{|{\bf r}-{\bf r}^\prime|^2}\cos\alpha ~d^3{\bf r}^\prime\,,
\end{eqnarray}
where $\cos\alpha$ is the projection along the line of centres and $\alpha$ is the angle indicated in Figure~\ref{spiral}.

Due to the interaction the with Machian strings of $M$, the direct strings are pulled in towards the line of centres. Consider, first, the limiting case of complete alignment, so that all the direct strings are aligned parallel to the line of centres when they join on to $m$. Putting $\alpha= 0$ in~(\ref{adirect}), the gravitational acceleration in the case of complete direct string alignment is
\begin{eqnarray}\label{adirect2}
a(r)\,=\,\int \frac{G\rho({\bf r}^\prime)}{|{\bf r}-{\bf r}^\prime|^2}\,d^3{\bf r}^\prime\,.
\end{eqnarray}
If the mass distribution of $M$ is spherically symmetric, so that $\rho({\bf r}^\prime)= \rho(r^\prime)$, then~(\ref{adirect2}) becomes
\begin{eqnarray}\label{adirect3}
a(r)&=&2\pi G \int \!\!\int\frac{{r^\prime}^2\rho(r^\prime)\,dr^\prime dcos\theta}{r^2 - 2rr^\prime\cos\theta + {r^\prime}^2}\nonumber\\
 &=& \frac{2\pi G}{r} \int \!\rho(r^\prime)\,r^\prime \ln\frac{r+r^\prime}{|r-r^\prime|}\,dr^\prime\,.
\end{eqnarray}
 In the limit when $M$ is a point mass, or when the distance $r$ between the two masses is much larger than the size of the mass distribution of $M$,~(\ref{adirect3}) reduces to the Newtonian gravitational acceleration $GM/r^2$. 
 
 It is of interest to investigate the effect of complete direct string alignment on galaxy rotation curves. If the galaxy is modelled as a two-dimensional exponential disc with scale height $h_d$, so that the surface mass density of the disc at a radius $r$ is proportional to $e^{-r/{h_d}}$, then it is a standard result~\cite{freeman} that the velocity rotation curve in Newtonian gravity is given by
\begin{eqnarray}\label{vrot}
v^2(r)\,=\,\frac{GM}{h_d}\frac{x^2}{2}\Big[I_0\Big(\frac{x}{2}\Big)K_0\Big(\frac{x}{2}\Big) - I_1\Big(\frac{x}{2}\Big)K_1\Big(\frac{x}{2}\Big)\Big]\,,\nonumber\\
\end{eqnarray}
where $M$ is the total mass of the disc, $x= r/h_d$ and $I$ and $K$ are Bessel functions. 

Now consider the gravitational acceleration due to an exponential disc in the aligned direct string model. The integral~(\ref{adirect2}) actually diverges for an infinitely thin two dimensional disc so the exponential disc is defined as a uniform volume density, $\rho$, for $0\le z\le h_0 e^{-r/{h_d}}$, where $z$ is the distance normal to the plane of the disc and $h_0$ is the central thickness. The total mass of the disc is then $M= 2\pi\rho h_0h_d^2$ so~(\ref{adirect2}) becomes, in cylindrical polar coordinates,
\begin{eqnarray}\label{adirect4}
a(r)=\frac{GM}{2\pi h_0 h_d^2}\int_0^\infty\!\!\!\!s\,ds\!\int_0^{2\pi} \hspace*{-0.3cm}d\theta\!\int_0^{h_0e^{-s/{h_d}}} \hspace*{-1cm}\frac{dz}{r^2-2rs\cos\theta + s^2 + z^2}.\hspace*{-0.5cm}\nonumber\\
\end{eqnarray}
The integral over $\theta$ may be performed using~\cite{grad}
\begin{eqnarray}\label{form}
\int_0^{2\pi}\!\!\frac{d\theta}{a+b\cos\theta}\,=\,\frac{2\pi}{a^2-b^2}
\end{eqnarray}
to give
\begin{eqnarray}\label{adirect5}
a(r)=\frac{GM}{h_0 h_d^2}\int_0^\infty\!\!\!\!s\,ds\!\int_0^{h_0e^{-s/{h_d}}} \hspace*{-1cm}\frac{dz}{\sqrt{(r^2+ s^2 + z^2)^2 - 4r^2s^2}}.\hspace*{-0.5cm}\nonumber\\
\end{eqnarray}
Interchanging the order of integration gives
\begin{eqnarray}\label{adirect6}
a(r)=\frac{GM}{h_0 h_d^2}\int_0^{h_0}\!\!dz \!\int_0^{h_d\ln(h_0/z)} \hspace*{-1cm}\frac{s\,ds}{\sqrt{(r^2+ s^2 + z^2)^2 - 4r^2s^2}}\hspace*{-0.5cm}\nonumber\\
\end{eqnarray}
and changing variables from $s$ to $x=s^2$ then gives
\begin{eqnarray}\label{adirect7}
a(r)=\frac{GM}{2h_0 h_d^2}\int_0^{h_0}\!\!dz \!\int_0^{[h_d\ln(h_0/z)]^2} \hspace*{-1cm}\frac{dx}{\sqrt{(x+r^2 + z^2)^2 - 4xr^2}}\,.\hspace*{-0.5cm}\nonumber\\
\end{eqnarray}
The inner integral may be evaluated using~\cite{grad2}
\begin{eqnarray}\label{form2}
\hspace*{-0.5cm}\int \frac{dx}{\sqrt{a + 2bx + x^2}}\,=\,\ln\Big(\frac{\sqrt{a+2bx+x^2} + x + b}{\sqrt{a-b^2}}\Big)\nonumber\\
\end{eqnarray}
to give
\begin{eqnarray}\label{adirect8}
a(r)\,=\,\frac{GM}{2h_0 h_d^2}\int_0^{h_0}\!dz~F(z)\,,
\end{eqnarray}
where the function $F(z)$ is defined by
\begin{eqnarray}\label{fdef}
F(z)\,=\,\ln\Big[\frac{\sqrt{(x_d+r^2 + z^2)^2 - 4x_dr^2} \,+ x_d + z^2-r^2}{2z^2}\Big] \hspace*{-0.5cm}\nonumber\\
\end{eqnarray}
and $x_d= [h_d\ln(h_0/z)]^2$. The corresponding rotational velocity, $v(r)$, is defined by $a(r)= v(r)^2/r$. The numerical integration of~(\ref{adirect8}) is simplified by the change of variables from $z$ to the dimensionless variable $y$ defined by $y= (h_d/h_0)\ln(h_0/z)$. The formula for $v(r)$ is then
\begin{eqnarray}\label{vrot2}
v(r)^2\,=\,\frac{GM}{h_d}\frac{h_0r}{2h_d^2}\int_0^\infty\!dy~G(y)
\end{eqnarray}
where $G(y)$ is given by
\begin{eqnarray}\label{gdef}
G(y)\,=\,\ln\Big[\frac{\sqrt{(y^2+{\tilde r}^2 + {\tilde z}^2)^2 - 4y^2{\tilde r}^2} \,+ y^2 + {\tilde z}^2-{\tilde r}^2}{2{\tilde z}^2}\Big]\,, \hspace*{-0.5cm}\nonumber\\
\end{eqnarray}
with ${\tilde r}= r/h_0$ and ${\tilde z}= z/h_0$.

\begin{figure}[h]
\vspace*{0.5cm}
\includegraphics[height=4cm,width=7cm]{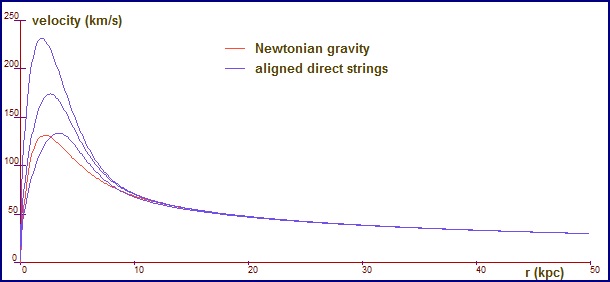}
\caption{\label{rotation} The rotation curve for an exponential disc of mass $M= 10^{10}M_{\odot}$ and scale height $h_d= 1\,$kpc in Newtonian gravity and in the aligned direct string model. The rotation curve in the aligned direct string model is shown for three values of the central disc thickness. The highest curve corresponds to $h_0 = 1\,$kpc, the middle curve corresponds to $h_0 = 6\,$kpc and the lowest curve corresponds to $h_0 = 20\,$kpc.}
\end{figure}

The functions~(\ref{vrot}) and~(\ref{vrot2}) are plotted in Figure~\ref{rotation} for an exponential disc of mass $M= 10^{10}M_{\odot}$ and scale height $h_d= 1\,$kpc, for different values of the central thickness, $h_0$. 
At large radii, the rotation curves are seen to be almost identical to the Newtonian rotation curve. The complete alignment of direct strings does has a significant effect at small radii, particularly for thin discs.

The assumption of complete direct string alignment, on which equation~(\ref{vrot2}) is based, is only an approximation. Indeed, the integral~(\ref{adirect2}) tends to a non-zero limit as $r\rightarrow 0$, whereas the actual acceleration must tend to zero at the origin by symmetry. Further work to calculate the precise configuration of direct strings is therefore needed. In the present paper, the requirement that the alignment of direct 
strings must tend to zero as $r\rightarrow 0$ is incorporated by hand, by first calculating the additional acceleration in the limiting case of complete direct alignment and then multiplying the resulting additional acceleration by a function of the form
\begin{eqnarray}\label{falign}
\frac{r}{r + r_d}\,,
\end{eqnarray}
for some length scale $r_d$, which tends to zero linearly as $r\rightarrow 0$ and tends to unity as $r\rightarrow \infty$. The length scale $r_d$ is expected to be the same order of magnitude as the characteristic size of the system under consideration, but is otherwise treated as a free parameter.

\vspace*{0.4cm}
 \section{Gravitational lensing}\label{applensing}

%The deflection of a photon from a background galaxy due to a given gravitational mass distribution is assumed to be twice the deflection predicted by Newtonian gravity, in accordance with experiment and with theories of gravity such as General Relativity and the model of Machian space quanta~\cite{paper1}. 

 The standard theory of gravitational lensing (see e.g.~\cite{peacock}) is summarised here for completeness. The derivation is based on the result that the gravitational deflection of a light ray passing at a distance $d$ from a mass $M$ is twice the Newtonian prediction, namely $4GM/dc^2$, and therefore applies equally well to the theory of Machian space quanta as to the standard theory based on General Relativity.

 Consider a source at a distance $D_s$ and a gravitational lens at a distance $D_l$. If the lens deflects a
 beam of light from the source through an angle $\widetilde{\alpha}$ then, by simple trigonometry, the image
 appears to be deflected through an angle $\alpha= D_{ls}\widetilde{\alpha}/D_s$, where $D_{ls}$ is the
 distance from the lens to the source. The apparent angular position of the source is therefore $\theta = \beta
 + \alpha$, where $\beta$ is the angular position of the source in the absence of the lens. The deflection angle
 is a function of the angle $\theta$ in the lens plane, so $\beta = \theta - \alpha(\theta)$. The gravitational deflection of a light ray passing at a distance $d$ from a mass $M$ is $\widetilde{\alpha}=4GM/dc^2$ so, for a lens with projected surface mass density $\Sigma$,
\begin{eqnarray}\label{alpha}
 {\pmb \alpha}({\bf x})\,=\,\frac{4GD_{ls}}{c^2D_s}\int \frac{({\bf x}-{\bf x}^\prime)\,\Sigma({\bf x}^\prime)}
 {|{\bf x}-{\bf x}^\prime|^2}d^2{\bf x}^\prime\,,
\end{eqnarray}
 where ${\bf x}$ is the two-dimensional position vector in the lens plane. The lens convergence, $\kappa$, is
 defined by
\begin{eqnarray}\label{kappa}
 \kappa({\pmb\theta})\,&\equiv&\,\frac{1}{2}\Big(\frac{\partial\alpha_x}{\partial\theta_x}+\frac{\partial
 \alpha_y}{\partial\theta_y}\Big)\nonumber\\&=&\,\frac{2GD}{c^2}\,{\pmb \nabla}^2\!\!\int\Sigma({\bf x}^\prime)
 \,\ln |{\bf x}-{\bf x}^\prime|\,d{\bf x}^\prime \nonumber\\&=&\,\frac{4\pi GD}{c^2}\,\Sigma({\bf
 x})\nonumber\\&\equiv&\,\Sigma/\Sigma_c\,,
\end{eqnarray}
 where $D\equiv D_{ls}D_l/D_s$ and the critical surface density, $\Sigma_c$, is
\begin{eqnarray}\label{sigc}
 \Sigma_c=\frac{c^2}{4\pi GD}\,\approx\,1.6\times 10^9 \,\Big(\frac{1\mbox{Gpc}}{D}\Big)\,M_{\odot}/
 \mbox{kpc}^2.~
\end{eqnarray}
 For the bullet cluster, $D\approx 540\,$kpc so $\Sigma_c\approx 3\times 10^9\,M_{\odot}$/kpc$^2$.

\end{appendix}

\bibliography{master}

\end{document}